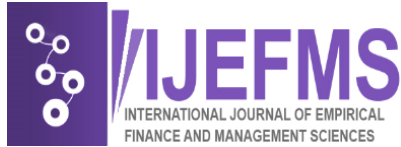

# Empirical Study to Explore the Influence of Salesperson's Customer Orientation on Customer Loyalty

**Author's Details:**
[1]**Prathamesh Muzumdar-**Research Associate-Dakota State University, Madison, South Dakota, USA
[2]**George Kurian-**Assistant Professor-Eastern New Mexico University, Portales, New Mexico

*Abstract*
*This study tries to examine the influence of salesperson's customer orientation on Customer Loyalty. Customer orientation is the approach taken by a salesperson to improve customer relationship and increase sales. Many organizations prefer sales orientation as a strategic approach towards increasing sales. Though successful in its objective, sales orientation fails to attract repetitive purchase. It has become a necessity to train frontline employees to better understand the customer needs, keeping in mind the firm's ultimate objective. This study examines the improvements, customer orientation can bring to increase repurchases thus leading to customer loyalty. The findings suggest that product assortment, long lines of customers, customers' annual income, and the listening skills of salesperson were the significant antecedents of customer loyalty.*
***Keywords:*** *Customer loyalty, Direct selling, Listening skills, Frontline employees, Customer orientation.*

## Introduction

Frontline employees play an important role in building relationship with customers. Their abilities to understand the need and interpret those needs play an important part in building long term relationships. The important aspect of two-way communication is listening which helps to build long term relationship (Ingram, Schwepker, & Hutson, 1992). Effective listening is influenced by various factors like customer orientation, adaptive selling, satisfaction, trust, salesperson performance, and intention to future interaction (Goad, 2014). Customer orientation (CO) is an approach which helps the employees to satisfy customer's long-term needs (Saxe & Weitz, 1982).

Customer orientation is a key component of market orientation and plays an important role in firm's performance (Kirca, Jayachandran, & Bearden, 2005). Customer orientation plays an integral part in developing the listening skills of an employee (Pelham, 2009). In general, customer orientation influences listening in a very positive way.

Customer loyalty is considered as a factor to strengthen the relationship between a customer's attitudes and repurchase (Dick & Basu, 1994). Attitudinal and behavioral loyalty are two measures of loyalty (Pan, Sheng, & Xie, 2012). Behavioral loyalty is largely affected by the interpersonal skills of a salesperson (Naumann, Widmier, & Jackson, 2000). Listening remains to be an integral part of salesperson's interpersonal skills (Ramsey & Sohi, 1997). This study tries to examine the effects of better listening on customer repurchase, as customer orientation improves a salesperson's listening ability, it indirectly influences customer loyalty in the form of product repurchase.

**Figure 1: Conceptual Framework**

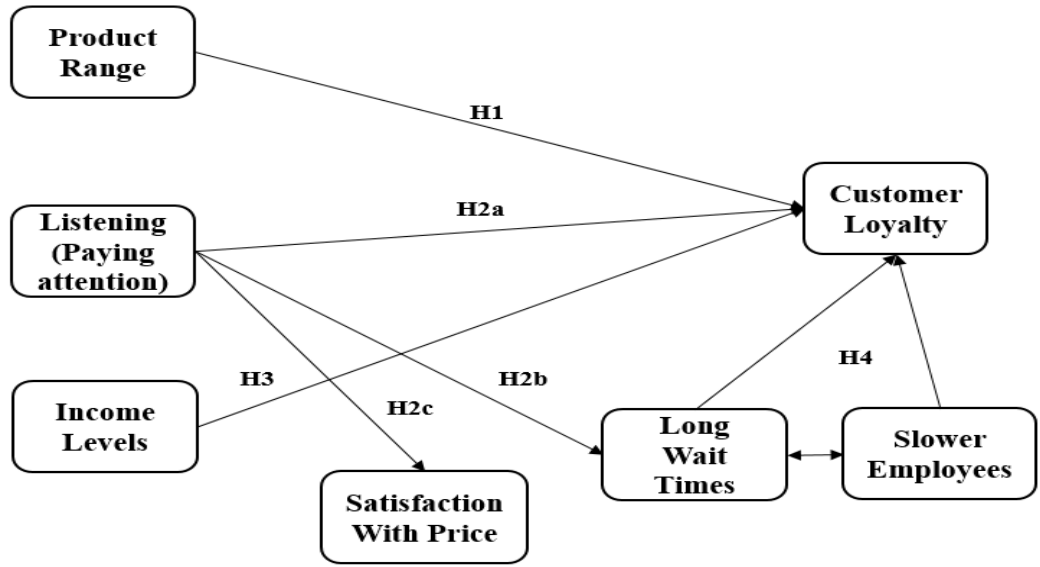

# Literature Review and Hypothesis

Previous researches indicate that there is a relationship between customer orientation and listening (Babakus, Cravens, Johnston, & Moncrief, 1999). Better customer orientation results in better listening eventually resulting in a satisfied customer (Bhuian, Bulent, & Borsboom, 2005). A salesperson's positive influence towards customer will result in customer's frequent visits to the retailer further resulting into repurchases (Boles, Johnston, & Hair, 1997).

## *Customer loyalty*

Customer loyalty is measured when a customer repurchases a product or products from the same retailer. It's a behavioral and attitudinal decision taken to address a person's need. Customer loyalty is the locus of customer retention and can be affected by multiple intrinsic and extrinsic factors, importantly, it is a sensitive pivot across which the fulcrum of loss and profit dwindles. This study's focal point is in understanding how customer loyalty can be influenced by the listening ability of a Frontline sales employee (Ingram, Schwepker, & Hutson, 1992). Throughout this study, the influence of other variables will be measured over customer loyalty.

## *Product Range*

List of available sets of products have influenced the purchase intention of customers for a long period of time (Castleberry & Shepherd, 1993). A long list of product range helps customers to make better choices in their selection and don't find themselves being pushed to purchase products (Comer & Drollinger, 1999). This leads to multiple repurchases further leading to loyalty towards a brand or the retailer. Having a long range of products influences the purchasing ability of the customer (Ramsey & Sohi, 1997). This study tries to find out the extent of influence long product range has on customer repurchase decision. Overall, the study tries to measure the influence of the product range on customer loyalty.

**H1:** Available long range of products positively affects customer repurchase decision in a product preference setup.

Hypothesis H1 tries to understand the influence long product range has on customer repurchase decision. The study tries to explore the relationship between the repurchase decision and more variety of products available at the retailer. The hypothesis aims to discover if the customer will be more willing to repurchase just on the factor of long range or do other factors influence this decision.

## *Listening salesperson*

Paying attention to customers is an important aspect of making a customer satisfied with your service. Customer are willing to purchase more if the salesperson is able to attend to their needs and provide them the service they look forward to (Saxe & Weitz, 1982). The ability of a salesperson to listen to customer needs and comprehending on that information to develop a sense of customer's purchase intention can help the salesperson gain the customer confidence. This study tries to understand the influence of listening to customer needs on their repurchase intentions.

**H2a:** Paying attention to customer requirements (sales people listening) positively influence a customer's loyalty.

Hypothesis H2a tries to understand the influence of salesperson's listening skills on customer's repurchase behavior. The study tries to explore if there exists a positive influence of paying attention to customer needs on product repurchase.

**H2b:** Paying attention to customer requirement suppresses the negative effects of long wait times.

The hypothesis H2b addresses how paying attention can suppress negative effects of the long wait time. Paying attention can generate better relationships leading to repurchase (Comer & Drollinger, 1999). Price plays an important in product purchase (Dixon & Schertzer, 2005). Customers get heavily influenced by product pricing and pricing sometimes determine the firm's position (Dubinsky & Ingram, 1984). It is important that a customer should be satisfied with the price he pays to the firms. Employee service plays an important role in customer satisfaction (Low, Cravens, Grant, & Moncrief, 2001). Paying attention to customer and serving the customer appropriately can lead to multiple purchase of price sensitive products (McFarland, 2003).

**H2c:** Paying attention to customer requirement increases a customer satisfaction towards the product price satisfaction.

The hypothesis H2a, H2b, and H2c address how paying attention positively affects a customer's repurchase intention. Do customer prefer Frontline employees to understand and interpret their needs and how often they feel that their needs are satisfied with the interpersonal listening skills of Frontline employee will be taken into consideration.

## *Income*

Customer income plays an important part in deciding whether the customer would be able to purchase the product or not (Pan, Sheng, & Xie, 2012). It is very important for the organization to understand the influence of customer income on customer's repurchase behavior. In this study the customer income is divided into five categories and they are tested on the customer loyalty.

**H3:** Upper and lower middle class and working class income level positively influence customer loyalty.

The hypothesis H3 helps to explore the influence of five categories of income on customer repurchase decision. It helps to understand which income category significantly influences the repurchase decision.

## *Long wait time, slower employees, and pushy salespeople*

Long wait time leads to long wait times. Slow employees become a bottleneck during peak purchase periods, where long wait times are inevitable. Few Frontline employees' result in long wait time for consumers to complete their orders. It is very important to understand how listening abilities can suppress the negative effects of long wait times (Brown & Peterson, 1993). Long wait time, usually results in customer dissatisfaction further resulting in abandoning repurchases (Castleberry & Shepherd, 1993). It is very important to understand how customer orientation can actually lead to better customer satisfaction to take into consideration the fewer Frontline employees (Chonko & Burnett, 1983).

**H4:** Long wait time and slower employees are both responsible for affecting customer loyalty.

The hypothesis H4 tries to examine how listening subsides the negative effects of price. Do consumers prefer to buy expensive product because they feel satisfied by the service provided to them by Frontline employees.

## Data Collection

The data collection was done using survey methodology. A sample consisting of a panel of consumers who interact with Frontline employees for purchase was selected from the responses to the survey screener. A quantitative questionnaire was designed in this phase. The questionnaire was divided into two parts, it contained a screener followed by the main section. The survey was administered using an online survey instrument. Data was analyzed using SPSS. Overall quantitative methods were used to analyze the data.

The data were collected through an online survey. The survey consisted of 56 questions and took an average of 62 minutes to complete. Customers from five different retail outlets belonging to same company were used for this study. A screener was placed before the main survey to filter the respondents, who did visit the store but didn't purchase. The basic filter criteria was to select respondents who at least made a single purchase at the retail outlet. The selected respondents were sent an email link to the survey. The survey was distributed among 4500 respondents and 2332 completed responses were collected with a response rate of 51%. Only completed surveys with each question answered were used for data analysis.

## Demographics

The survey consisted of a section which recorded demographic information of the respondents such as age, gender, marital status of the customer. The demographics recorded report that 66% were females and 34% were male. 40% of the respondent were full-time employed, 22% were part-time employed, 6% were unemployed, 20% were students and not employed, 12% percent were retired. 20% of respondents were in the age group of 18 to 25 years, 42% were in the age group of 26 to 35 years, 22% were in the age group of 36 to 50 years, and 16% were 51 years and above.

## Research Methodology

### Logistic Regression and ANOVA

The causal research design was used to analyze the data. Casual research was also used to test the hypothesis. The cause and effect relationship was determined by manipulating the independent variable to check its effect on the dependent variable while controlling the effect of extraneous variables. Logistic regression was used to understand the causal relationship between the dependent and independent variables. Logistic regression is used for the methodological approach because it allows the dependent variable to be categorical in nature. Other methods like linear regression which has similar predictability can predict the causal relationship, though they don't the dependent variable to be categorical in nature.

It was used to test **hypothesis H1, H2a, and H3**. Initially ANOVA was used to explore the significant variables. It was found that five out of seven variables are significant and can be used for testing causal relationship in logistic regression. The five variables were considered in the logistic regression where the variable of income was also taken into consideration. The dependent variable in the regression was loyalty which is categorical in nature. The independent variable considered were these five variables, including five levels (dummy variable) for the variable income. The main purpose to include the variable income at this stage was to analyze the effect of each income category on the customer loyalty, the intention to repurchase again.

## Pearson's Correlation Coefficient

Pearson's correlation coefficient was used to test **hypothesis H2b, H2c, and H4** in conjunction with the logistic regression results. Variables long wait time and slower employees being the interval in nature were analyzed using Pearson's correlation to explore the significance and direction of the relationship. Further that result was used in conjunction with the results from logistic regression to verify the overall significance in relationship with the dependent variable customer loyalty, which is nominal in nature.

# Results

### Table 1: Result of the ANOVA

| Variable | F- values | Significance level |
|---|---|---|
| Product Range | 7.11** | 0.00 |
| Long Wait Time | 39.47** | 0.00 |
| Slower Employees | 30.80** | 0.00 |
| Pushy Salespeople | 0.55 | 0.45 |
| Disorganized Store | 2.56 | 0.11 |
| Listening salesperson | 41.11** | 0.00 |
| Satisfied with prices | 4.26 | 0.42 |

Note: Variables are tested at 1% (0.01) significance level

In the logistic regression, ANOVA was initially used to find out which variables differ over the groups. It was found (Table 1) that out of seven variables, four were found to be highly significant (0.000%). The F ratio of all the four variables was also found to be very high. Out of the four variables, sales people listen to me, the attribute of paying attention had the highest F ratio (41.11) followed by long wait time (queue) (39.47), and slow employees (30.8). Other three variables were found non-significant even at 10% significance level.

### Table 2: Result of Logistic Regression

|  | Beta | Standard Error | Wald | df | Significance Level | Exp(B) |
|---|---|---|---|---|---|---|
| Product Range | 0.85 | 0.27 | 9.44 | 1 | 0.00 | 2.34 |
| Long wait times | -0.72 | 0.18 | 15.22 | 1 | 0.00 | 0.48 |
| Listening Salesperson | 0.66 | 0.21 | 9.39 | 1 | 0.00 | 1.94 |
| Slower Employees | -0.76 | 0.35 | 9.62 | 1 | 0.00 | 3.31 |
| Income (Below $10,000) | 2.41 | 1.67 | 2.08 | 1 | 0.15 | 11.21 |
| Income (10,001 to 25,000) | 3.64 | 1.49 | 5.98 | 1 | 0.01 | 38.38 |
| Income (25,001 to 50,000) | 4.14 | 1.51 | 7.43 | 1 | 0.00 | 62.78 |
| Income (50,001 to 75,000) | 3.39 | 1.37 | 6.06 | 1 | 0.01 | 29.74 |
| Income (75,001 and above) | 2.72 | 1.55 | 3.05 | 1 | 0.08 | 15.17 |
| Constant | -4.71 | 1.78 | 6.97 | 1 | 0.00 | 0.009 |
| Chi Square | 68.67 | | | | | |
| Sig | 0.00 | | | | | |
| -2 Log likelihood | 60.81[a] | | | | | |
| Cox & Snell R Square | 0.5 | | | | | |
| Nagelkerke R Square | 0.684 | | | | | |
| Hit Ratio | 88[b] | | | | | |
| Base model Hit Ratio | 65[b] | | | | | |

Note: Variables are tested at 1% (0.01) significance level

From table 2 it is clear that three variables out of four, which were derived from the ANOVA are found to be significant. Product range (0.002), long wait time (0.000), and the listening salesperson (0.002) were all found to be significant. In the income variable level, three income levels were found to be significant. Income (10,001 to 25,000), (25,001 to 50,000), and (50,001 to 75,000) were found to be highly significant, whereas level 1 and level 5 were non-significant. It was found that product range (2.34) has the highest odds ratio followed by listening salesperson (1.94) and long wait time (0.48). The whole equation was found to be highly significant with high chi square and a hit ratio of 88.

**Table 3: Result of Pearson Correlation Matrix**

|  | Product Range | Long wait times | Slower employees | Pushy salespeople | Disorganized store | Listening Salesperson | Satisfied with prices |
|---|---|---|---|---|---|---|---|
| Product Range | 1.0 | 0.13 | 0.91 | -0.04 | 0.17 | 0.1 | 0.01 |
| Long wait times | - | 1.0 | 0.69** | 0.2 | 0.16 | -0.4** | -0.16 |
| Slower employees | - | - | 1.0 | 0.25 | 0.06 | -0.61** | -0.17 |
| Pushy salespeople | - | - | - | 1.0 | 0.24* | -0.1 | -0.1 |
| Disorganized store | - | - | - | - | 1.0 | -0.12 | -0.48** |
| Listening Salesperson | - | - | - | - | - | 1.0 | 0.2** |
| Satisfied with prices | - | - | - | - | - | - | 1.0 |

** Correlation is significant at the 0.01 level (2-tailed)
 * Correlation is significant at the 0.05 level (2-tailed)

Pearson's correlation coefficient was used to test hypothesis H2b, H2c, and H4. From Pearson's correlation it was found that long wait time is significantly positively correlated (0.69) with slower employees and at the same time negatively correlated (-0.4) with listening salesperson. From the Logistic regression it is confirmed that variables long wait time and slower employees, both significantly affect the customer loyalty.

Listening salesperson is significantly positively correlated (0.2) with the variable satisfied with the prices. This helps us to infer that customer satisfaction with a salesperson can compromise towards the satisfaction with the price.

## Discussion

From the analysis, we conclude that we fail to reject hypothesis H1. The product range has the highest odds ratio, making it very certain that long range of cell phones on display help consumers more on deciding the product rather than other aspects of the sales. The product range is seen to be the biggest advantage of retail outlets. We fail to reject Hypothesis H2a as listening salesperson or the attribute of paying attention to the consumer is seen to have the second highest impact of the repurchase decision of a consumer. Customer's loyalty towards a particular outlet is strongly seen to be affected by it, visiting the same outlet and purchasing a product through the same agent is seen to be the preferred method of purchase.

We fail to reject Hypothesis H3 where three income level income (10,001 to 25,000), (25,001 to 50,000), and (50,001 to 75,000), which denote upper middle class, lower middle class, and working class have odds ratios which are significant (0.01). Lower middle (25,001 to 50,000) class income level is seen to have the highest influence on the repurchase decision, where it is precisely from the odds ratio (62.78) that customer turnout in the income (25,001 to 50,000) would be the highest.

From the analysis, it is clear from the results there exists a negative correlation between paying attention to customer requirement and long purchase lines and employees' efficiency towards a sale. The more the salesperson is able to accommodate a customers' views, the more are the chances a customer will be retained for repurchase, thus we fail to reject hypothesis H2b.

This also implies that a customers' discomfort from long purchase line and tardiness of the salesperson can be suppressed by a salesperson's ability to please the customer through his communication and listening skills. Salespersons' paying attention ability is also seen to have a positive correlation with a customer satisfaction with product prices. This correlation actually benefits in a way that it helps a salesperson to curb the price negotiation to a certain point (price) therefore we fail to reject hypothesis H2c.

From the analysis, we fail to reject hypothesis H4, as there is a significant positive correlation between long wait times and slower employees. This implies that slower employees can lead to inefficient operations leading to long wait times. From the logistic regression results, it is clear that both the variables are significantly correlated to customer loyalty. Thus we conclude that long wait times and slower employees decrease the repurchase intent further negatively influencing customer loyalty.

## Conclusion

This study helps us to understand how a retail organization builds strategies along there Frontline employees to improve their sales through customer loyalty. This study proves that product range and income levels influence the customer loyalty. Better product range can give the customer satisfaction of choice and choosing products from wider assortment. The customer won't feel being pushed over to make choices over a confined range. This study fulfills the aim of proving that listening ability to form of paying attention to customer needs can help improve customer loyalty. This helps us to better understand the influence of customer orientation and salesperson training over customer retention. It is also proved in this study that paying attention to customers can also bring more satisfaction among customers about the product price.

Eventually the study helps us to understand that slower employees can create long wait time further negatively influencing customer loyalty. Training employees to increase efficiency by focusing on the needs of the customers and effectively reducing the wait time can help retain a customer.

Organizations highly depend upon salesperson- consumer interaction for sales performances. This study will help firms to improve their employee training and make their organizations more customer oriented. The results indicate if organizations can allocate sufficient amount of budget to train Frontline sales personal they can improvise on skill improvements of the employees. Listening to the needs of the customer is one of the strongest skill set that can lead to a better customer relationship, further leading to better higher chances of repurchase. This experience can add up to developing higher loyalty in customers and improves their repurchase behavior. This can also make the Frontline employees more efficient in processing customer needs and improvise the stores operational and logistical efficiencies.

This study also helps organizations to understand the relevant income groups, which are responsible for the repurchase behavior and increased customer loyalty. Promotional efforts can be focused on certain groups who exhibit a higher tendency towards repurchase thus further influencing the loyalty. Product process and discounts can be designed towards certain income groups. Overall, this study helps organizations to better allocate budget and efforts towards effective strategies.